\documentclass[aps,prl,twocolumn,groupedaddress,showpacs]{revtex4}
\usepackage{graphicx}
\begin{document}
\title{p-spin model in finite dimensions and its relation
 to structural glasses  }
\author{M.A.Moore }
\affiliation{Department of Physics and Astronomy, University of 
Manchester,
Manchester, M13 9PL, United Kingdom}

\author{Barbara Drossel} 
\affiliation{Physics Department, Technical University of Munich,
85747 Garching, Germany}
\date{\today}
\begin{abstract}
The p-spin spin-glass model has been studied extensively at mean-field
level because of the insights which it provides into the  mode-coupling
approach to structural glasses and the nature of the glass transition. 
We demonstrate explicitly that the \textit{finite dimensional} version of the 
p-spin model is in the same universality class as an Ising spin glass in
a magnetic field. Assuming that the droplet picture of Ising spin
glasses is valid we discuss how this universality may provide insights
into why structural glasses are either ``fragile" or ``strong".
\end{abstract}
\maketitle

In recent years  structural glasses
have been studied by many researchers  using ideas from spin glasses.
This activity was initiated by Kirkpatrick, Thirumalai and Wolynes \cite{Kirk}
who observed the similarity between the dynamical behavior of p-spin spin
glass models at mean-field level and the mode-coupling equation for glasses 
\cite{Gotze}. At mean-field level, the p-spin model has two transitions \cite
{Bouchaud} of interest for the structural glass analogy. There
is a dynamical transition temperature $T_d$ below which the spin-spin 
correlation functions do not decay to zero in the long time limit which 
implies
a breakdown of ergodicity. It is this transition which corresponds to the
transition in the mode-coupling equations. The  transition at
$T_c<T_d$ is a transition associated with thermodynamic singularities (there
are none at the dynamical transition $T_d$) and
 below it
replica symmetry is broken at the ``one-step'' level \cite{Bouchaud}.
It is this transition which is usually identified 
with the glass transition (if any) of structural glasses and $T_c$ associated
with, say, the Kauzmann temperature $T_K$ \cite{Kauzmann}. 

It has long been recognised that the dynamical transition at $T_d$ is an
artifact of the mean-field limit \cite{Parisi},\cite{Drossel}.
The transition is due to the exponentially 
large number of states which trap the system for exponentially long times
forbidding the system in the thermodynamic limit from reaching equilibrium.
However, in finite dimensional systems activation processes over
finite free-energy barriers or nucleation processes will take place and
restore ergodicity, a feature which was not incorporated either into
the early versions of mode-coupling theory. Thus the type of transition
which occurs at $T_c$ is the only possible transition which might exist
in finite-dimensional p-spin models. In fact 
in this paper we will argue that this transition does not
survive either in finite dimensional models of p-spin glasses. If the 
connection between real glasses and p-spin models goes beyond the
mean-field limit then one would expect that there will be 
no genuine transition in real glasses either.

Our approach will be to start from a commonly used finite dimensional version
of the p-spin model \cite{Parisi} and study it using field-theoretic
 methods. 
We find that the field theory of the p-spin model (at least when
 p is three, the only
value that we shall consider here)
 is in the same universality
class as the Ising spin-glass in a magnetic field. The possibility
of such a connection was suggested  
by Parisi et al. \cite{Parisi} based on their  numerical
simulations and by Drossel et al. \cite{Drossel} from their  Migdal-Kadanoff
renormalisation group studies, but it has never to our 
knowledge been explicitly demonstrated. While both models lack time reversal
 invariance, at
mean-field level the p-spin model and the Ising spin glass in a magnetic 
field behave quite differently so that their similarity for
finite dimensional systems cannot be taken as obvious. This is because 
the phase below $T_c$ in  the mean-field
p-spin model has replica symmetry broken at the one-step level
\cite{Bouchaud}, whereas the transition in the mean-field 
Ising spin-glass in a magnetic field -- the transition discovered by
de Almeida and Thouless (AT)  \cite{AT,BR} --  has the
replica symmetry broken at all steps.

Away from the mean-field limit the existence of a transition for the
Ising spin glass in a magnetic field remains controversial. According
to the droplet picture \cite{BM,FH,McM}, the presence of
a field removes the phase transition just as a field applied to a
ferromagnet does.  However, advocates of the replica symmetry breaking
picture of spin glasses claim that the AT transition exists in finite
dimensions \cite{Marinari}. In what follows we shall interpret our
central result of the equivalence of the p-spin model with the Ising
spin glass in a field according to the droplet model, and not discuss
the consequences of this equivalence in the replica symmetry breaking
picture. To support this interpretation we have determined the
correlation length $\xi$ of the p-spin model as a function of the
temperature $T$ using the Migdal-Kadanoff approximation (MKA)
\cite{Drossel}. The correlation length rises rapidly below a certain
temperature before saturating at a large value, which is around 30
lattice spacings in three dimensions, but there is no evidence within
this approach of an actual phase transition and a diverging
correlation length.

In the three-spin model used for the present study, each site is
occupied by two Ising spins, $\sigma_i$ and $\tau_i$, and the
Hamiltonian is
\begin{eqnarray}
{\mathcal H}(\{\sigma\}, \{\tau\})= &-&\sum_{<ij>}\left( J_{ij}^{(1)}
\sigma_i\tau_i\sigma_j +J_{ij}^{(2)}
\sigma_i\tau_i\tau_j \right. \nonumber\\
&&{}+\left.J_{ij}^{(3)}\sigma_i\sigma_j\tau_j + 
J_{ij}^{(4)}\tau_i\sigma_j\tau_j\right), 
\end{eqnarray}
where ${<ij>}$ are nearest-neighbor pairs, and the couplings $J_{ij}^{(n)}$
 are 
chosen independently from a Gaussian distribution with zero mean and width $J$.

When the signs of all the spins are reversed, the sign of the Hamiltonian
changes, indicating the violation of time reversal symmetry. The thermal
averages $\langle\sigma_i\rangle$ and $\langle\tau_i\rangle$
are non-zero at \textit{all} finite temperatures. 

In order to derive a field theory for the three-spin model we use the
usual replica identity for the free energy $F$
\begin{equation}
F=-kT\overline{{\rm ln}(Z)}=-kT\lim_{n\rightarrow 0}\frac{\overline{Z^n}-1}{n}
\end{equation}
to perform the average over the couplings $J_{ij}^{(n)}$, indicated
by the overline. To evaluate $\overline{Z^n}$ we set up 
$n$ replicas of the partition function $Z$
\begin{equation}
Z^n={\rm Tr}\prod_{\alpha=1}^{n}\exp
\left[-{\mathcal H}\left(\{\sigma^\alpha\},\{\tau^\alpha\}\right)/kT \right].
\end{equation}
${\rm Tr}$ denotes taking the sum over the values $\pm 1$ of
the Ising spins $\sigma_i^{\alpha}$ and $\tau_i^{\alpha}$ at each site.
After bond averaging one gets
\begin{eqnarray}
\overline{Z^n}={\rm Tr}\,\exp\bigg[\beta^2J^2nzN+
\frac{1}{2}\beta^2J^2\sum_{<ij>}
\sigma_i^{\alpha}\sigma_i^{\beta}\tau_i^{\alpha}\tau_i^{\beta}
\nonumber\\\times \left(\sigma_j^{\alpha}\sigma_j^{\beta}+\tau_j^{\alpha}\tau_j^{\beta}\right)
+\sigma_j^{\alpha}\sigma_j^{\beta}\tau_j^{\alpha}\tau_j^{\beta}
\left(\sigma_i^{\alpha}\sigma_i^{\beta}+\tau_i^{\alpha}\tau_i^{\beta}\right)\bigg],
\end{eqnarray}
where $z$ is the number of neighbors of a site (six for a simple cubic 
lattice) and $N$ is the number of lattice sites.

The terms involving different sites are decoupled by introducing fields
$q_{ij}^{\alpha\beta}$ and $p_{ij}^{\alpha\beta}$ for each  bond
$<ij>$ using:
\begin{eqnarray}
\lefteqn{\overline{Z^n}=
{\rm Tr}\,\exp\left(\beta^2J^2nzN\right)
\prod_{<ij>}^{}\frac{2}{\pi^{2}\beta^{4}J^{4}}
\int d^2q_{ij}^{\alpha\beta}
\int d^2p_{ij}^{\alpha\beta}}\nonumber\\
&&\times\exp\bigg[-\frac{\sqrt{2}}{\beta^2J^2}\left(q_{ij}^{\alpha\beta}
q_{ij}^{\alpha\beta\ast}+p_{ij}^{\alpha\beta}
p_{ij}^{\alpha\beta\ast}\right)\nonumber \\
&&{}+q_{ij}^{\alpha\beta}\sigma_i^{\alpha}\sigma_i^{\beta}
\tau_i^{\alpha}
\tau_i^{\beta}+{p_{ij}^{\alpha\beta}}\left(\sigma_i^{\alpha}\sigma_i^{\beta}
+\tau_i^{\alpha}\tau_i^{\beta}\right)/\sqrt{2}\nonumber\\
&&{}+p_{ij}^{\alpha\beta\ast}\sigma_j^{\alpha}\sigma_j^{\beta}
\tau_j^{\alpha}
\tau_j^{\beta}+{q_{ij}^{\alpha\beta\ast}}\left(\sigma_j^{\alpha}\sigma_j^{\beta}
+\tau_j^{\alpha}\tau_j^{\beta}\right)/\sqrt{2}\bigg],
\end{eqnarray}
where in the applied summation convention used here 
and in what follows for  terms containing
 $\alpha$ and $\beta$
those for which $\alpha=\beta$ are to be omitted. We next follow
 the procedure of 
Pytte and Rudnick \cite{Pytte} and define 
\begin{eqnarray}
&Q_{\alpha\beta}(i)\equiv\sum_{j}q_{ij}^{\alpha\beta}\,\nonumber\\
&P_{\alpha\beta}(i)\equiv\sum_{j}p_{ij}^{\alpha\beta}.
\end{eqnarray}
We are left with the following types of sum at each site:
\begin{equation}
{\rm Tr}\,\exp\left[Q_{\alpha\beta}(i)\sigma_i^{\alpha}\sigma_i^{\beta}
\tau_i^{\alpha}
\tau_i^{\beta}+\frac{P_{\alpha\beta}^{\ast}(i)}{\sqrt{2}}\left(\sigma_i^{\alpha}\sigma_i^{\beta}
+\tau_i^{\alpha}\tau_i^{\beta}\right)\right].
\end{equation}
These can be done by expanding the exponential, taking the trace and 
then re-exponentiating the result. Repeating this procedure
for each site and finally taking the continuum limit one
obtains, correct to $O(Q_{\alpha\beta}^3,P_{\alpha\beta}^3)$,
the following free-energy density
\begin{eqnarray}
\mathcal{F}&=&\frac{1}{4}{\nabla}Q_{\alpha\beta}{\nabla}
Q_{\alpha\beta}^{\ast}
+ \frac{1}{4}{\nabla}P_{\alpha\beta}{\nabla}P_{\alpha\beta}^{\ast}
+r\big(Q_{\alpha\beta}Q_{\alpha\beta}^{\ast}\nonumber\\
&&
{}+P_{\alpha\beta}P_{\alpha\beta}^{\ast}\big) 
-a(Q_{\alpha\beta}^2 +Q_{\alpha\beta}^{\ast2}
+P_{\alpha\beta}^2+ P_{\alpha\beta}^{\ast2})\nonumber\\
&-&\frac{1}{6}\tilde{w}_1\left(Q_{\alpha\beta}Q_{\beta\gamma}Q_{\gamma\alpha}
+Q_{\alpha\beta}^{\ast}
Q_{\beta\gamma}^{\ast}Q_{\gamma\alpha}^{\ast}\sqrt{2}\right)\nonumber\\
&-&\frac{1}{6}\tilde{w}_1\left(P_{\alpha\beta}
P_{\beta\gamma}P_{\gamma\alpha}/\sqrt{2}
+ P_{\alpha\beta}^{\ast}P_{\beta\gamma}^{\ast}P_{\gamma\alpha}^{\ast}\right)
\nonumber\\
&-&\frac{1}{6}\tilde{w}_2\left(P_{\alpha\beta}^2Q_{\alpha\beta}
+P_{\alpha\beta}^{\ast}
Q_{\alpha\beta}^{\ast2}\right),
\label{full}
\end{eqnarray}
where all fields have the symmetry $Q_{\alpha\beta}\equiv Q_{\beta\alpha}$,
$\alpha\ne\beta$ and the implied summation convention is used. Notice that
the free energy derived from the functional $\mathcal{F}$ is real as 
can be shown by taking the complex conjugate of Eq.(\ref{full}) and making
a variable change $P\Leftrightarrow Q$ in the fields, which are of course
integrated over to get the free energy $F$.

The above field theory, while complicated, is of standard type. The field
theory for the p-spin model obtained by Campellone et al. \cite{Campellone}
is very unusual and does not lend itself to a  
diagrammatic expansion. Truncating the expansion at the cubic terms is
adequate provided that the phase transition, if any, is continuous.
At mean-field level, the transition is continuous provided $w_2/w_1$ 
is sufficiently small (see below). The numerical studies of the
three-spin model
in finite dimensions suggest that the transition, if it exists, is
continuous, so that we shall focus on situations in which the 
truncation to a cubic field theory is sufficient. When this is the case
further simplifications are possible by simplifying the field
theory to the soft modes associated with the possible
transition. 

First of all, it is easy to see from the 
quadratic terms that as the coefficient $a$ is positive
the terms involving the imaginary parts of $Q$ and $P$ are associated with hard modes
and so can all be dropped near the putative transition. The  associated free 
energy  density functional is then
\begin{eqnarray}
\mathcal{F}&=&\frac{1}{4}({\nabla}A_{\alpha\beta})^2 
         +\frac{1}{4}({\nabla}B_{\alpha\beta})^2
-\frac{1}{4}t(A_{\alpha\beta}^2+B_{\alpha\beta}^2) \nonumber\\
&-&\frac{1}{6}w_1(A_{\alpha\beta}A_{\beta\gamma}A_{\gamma\alpha}+
B_{\alpha\beta}B_{\beta\gamma}B_{\gamma\alpha}) \nonumber\\
&-&\frac{1}{6}w_2(A_{\alpha\beta}^2B_{\alpha\beta}+A_{\alpha\beta}
B_{\alpha\beta}^2),
\end{eqnarray}
where $A_{\alpha\beta}$ is the real part of $Q_{\alpha\beta}$ and 
$B_{\alpha\beta}$ is the real part of $P_{\alpha\beta}$; 
$-t/4=(r-2a)$, $w_1=(1+1/\sqrt{2})\tilde{w}_1$, $w_2=\tilde{w}_2$.
This functional is manifestly real. 

A second simplification comes from observing that even at high temperatures
$\langle A_{\alpha\beta}\rangle =\langle B_{\alpha\beta}\rangle \ne0$, that is,
 the average
 of the
fields is always non-zero. This is due to the term 
$(A_{\alpha\beta}^2B_{\alpha\beta}+A_{\alpha\beta}B_{\alpha\beta}^2)$.  Any
transition which occurs will involve a breaking of this 
\textit{replica symmetric} solution. It is convenient to first introduce the
 new fields
\begin{eqnarray}
C_{\alpha\beta}&=&(A_{\alpha\beta}+B_{\alpha\beta})/\sqrt{2} \nonumber\\
D_{\alpha\beta}&=&(A_{\alpha\beta}-B_{\alpha\beta})/\sqrt{2}.
\end{eqnarray}
Then in terms of these new fields the free energy functional becomes
\begin{eqnarray}
\mathcal{F}&=&\frac{1}{4}({\nabla}C_{\alpha\beta})^2 +
          \frac{1}{4}({\nabla}D_{\alpha\beta})^2 -
          \frac{1}{4}t(C_{\alpha\beta}^2+D_{\alpha\beta}^2) \nonumber\\
&-&\frac{1}{6}v_2(C_{\alpha\beta}^3-C_{\alpha\beta}D_{\alpha\beta}^2)
-\frac{1}{6}v_1(C_{\alpha\beta}C_{\beta\gamma}C_{\gamma\alpha}\\
&+&D_{\alpha\beta}D_{\beta\gamma}C_{\gamma\alpha}+
D_{\alpha\beta}C_{\beta\gamma}D_{\gamma\alpha}
+C_{\alpha\beta}D_{\beta\gamma}D_{\gamma\alpha}), \nonumber
\end{eqnarray}
where $v_1=w_1/\sqrt{2}$ and $v_2=w_2/\sqrt{2}$.
If one introduces a field $s_{\alpha\beta}$ such that $C_{\alpha\beta}= 
s_{\alpha\beta}+C$ with $<s_{\alpha\beta}>=0$ then one can see 
that the terms involving the $D_{\alpha\beta}$ fields
are hard modes and can therefore be dropped. The free
energy density functional is then
\begin{equation}
\mathcal{F}=\frac{1}{4}({\nabla}C_{\alpha\beta})^2  
-\frac{1}{4}tC_{\alpha\beta}^2 -\frac{1}{6}v_1C_{\alpha\beta}C_{\beta\gamma}C_{\gamma\alpha}
-\frac{1}{6}v_2C_{\alpha\beta}^3.
\label{end}
\end{equation}
Ferrero et al.~\cite{Ferrero} 
wrote down this functional as the generic functional for 
discussing replica symmetry breaking at the one step level.
They showed that the transition at mean-field level is continuous
when $v_2/v_1\le1$. 

The functional in terms of the $s_{\alpha\beta}$ fields (leaving out
terms independent of $s_{\alpha\beta}$) is
\begin{eqnarray}
\mathcal{F}&=& \frac{1}{4}({\nabla}s_{\alpha\beta})^2
-\frac{1}{2}(t+v_1C^2+v_2C^2)s_{\alpha\beta}\nonumber\\
&-&\frac{1}{4}(t+2v_2C)s_{\alpha\beta}^2
-\frac{1}{2}v_1Cs_{\alpha\beta}s_{\alpha\gamma}\nonumber\\
&-&\frac{1}{6}v_1s_{\alpha\beta}s_{\beta\gamma}s_{\gamma\alpha}
-\frac{1}{6}v_2s_{\alpha\beta}^3,
\end{eqnarray}
where in the term $\frac{1}{2}v_1Cs_{\alpha\beta}s_{\alpha\gamma}$,
$\beta\ne\gamma$. Writing the quadratic terms in the form
\begin{equation}
\frac{1}{4}({\nabla}s_{\alpha\beta})^2
+M_{\alpha\beta,\gamma\delta}s_{\alpha\beta}s_{\gamma\delta},
\end{equation}
it turns out that the matrix $\mathbf{M}$ has three distinct
 eigenvalues \cite{AT}.
The limit of stability of the replica symmetric high-temperature phase
at mean-field level is determined by the vanishing of the smallest of these. 
AT showed that the corresponding eigenvectors span an $\frac{1}{2}n(n-3)$
dimensional subspace of the $\frac{1}{2}n(n-1)$ dimensional space 
defined by the $s_{\alpha\beta}$. This is called the ``replicon subspace"
\cite{BMR} and since only the replicon modes go soft at the AT or
replica symmetry breaking transition it suffices again to retain only
these modes for a study of the critical behavior of this transition.
It is useful to introduce the projection operator 
$S_{\alpha\beta,\gamma\delta}$ which projects any field $s_{\alpha\beta}$
onto the replicon subspace \cite{BR}.  Then
$\tilde{q}_{\alpha\beta}=\sum S_{\alpha\beta,\gamma\delta}s_{\gamma\delta}$,
where the sum is over distinct pairs $(\gamma\delta)$. The matrix elements are
$S_{\alpha\beta,\alpha\beta}=3, S_{\alpha\beta,\alpha\gamma}=3/2,
 (\beta\ne\gamma)$
and $S_{\alpha\beta,\gamma\delta}=1, (\alpha,\beta,\gamma,\delta$ all 
different) when $n$ is set to zero. Then our final free energy
density functional is
\begin{equation}
\mathcal{F}=\frac{1}{4}({\nabla}\tilde{q}_{\alpha\beta})^2
+\frac{1}{4}\tilde{t}\tilde{q}_{\alpha\beta}^2
-\frac{1}{6}v_1\tilde{q}_{\alpha\beta}\tilde{q}_{\beta\gamma}
\tilde{q}_{\gamma\alpha}
-\frac{1}{6}v_2\tilde{q}_{\alpha\beta}^3,
\label{BRE}
\end{equation}
where the coefficient $\tilde{t}$ is a measure of the distance
from the transition (if any).
Bray and Roberts \cite{BR} first obtained
this functional  in  their study of
the critical exponents in dimension $6-\epsilon$ at the AT line.

Thus the
ultimate field theory for the transition (if any) in 
the p-spin model is that of the Ising spin glass in a magnetic field. 
Bray and Roberts \cite{BR} obtained  the 
recursion equations for the coupling constants  $v_1$ and $v_2$ 
correct to order $\epsilon$ 
but did not find any stable fixed point.
They suggested that a possible explanation for this might 
be that there is no transition. We
next provide some evidence using the Migdal-Kadanoff  
renormalisation group procedure that this indeed may be the correct 
interpretation of their perturbative renormalisation group calculation.

\begin{figure}
\includegraphics*[width=6.2cm]{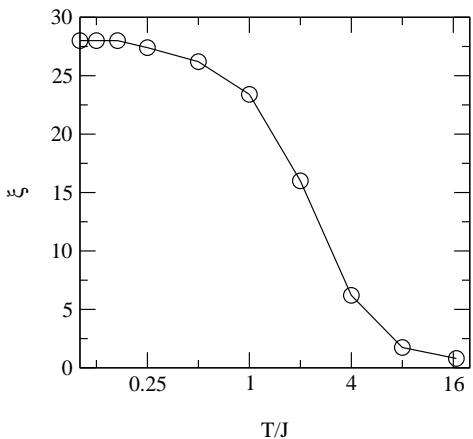}
\caption{The correlation length of the three-spin model in three
dimensions as function of $T/J$, as obtained using the MKA. \label{fig1}}
\end{figure}

The MKA is a real-space renormalization method which we have previously
applied \cite{Drossel} to both the three-spin model and the 
Ising spin glass in a field. For both models the behavior of
the couplings such as $J_{ij}^{(n)}$ under the renormalisation group was to 
flow to the high-temperature sink i.e.~zero, implying that there was no 
transition. We have now investigated how the effective couplings
decrease as a function of the
lengthscale $L$ and find that the variance of $J_{ij}^{(n)}$ decreases as
$\exp(-L/\xi)$, where $\xi(T)$ is the correlation length 
of the system.
The dependence of $\xi(T)$ on temperature is shown in Figure \ref{fig1} and it is
obvious that because of the rapid rise of $\xi(T)$ when $T\approx2J$ 
 Monte Carlo numerical studies on systems of
linear dimensions less than 10 could easily appear to have a 
phase transition \cite{Parisi}. 

What can we deduce from these results on p-spin glasses for
the physics of  structural glasses? The connection of the transition
at $T_c$ in the p-spin model (if any)  with the structural glass 
transition at around
$T_K$ (if any) has been much discussed \cite{Bouchaud}. 
The relationship can really only go beyond an analogy if the
two systems belong to the 
same universality class like, for example, the Ising ferromagnet and the 
liquid-gas critical point transition. At first sight any relationship 
between the p-spin model and phenomena in supercooled liquids seems unlikely.
The p-spin model has quenched disorder, and there is no disorder at all
in the glass transition problem. If there is a connection 
it would probably 
arise from similarities of the pure states in the two systems. The one-step
replica symmetry broken state below $T_c$ implies that there are 
many pure states present each with the same overlap with each other 
\cite{Bouchaud}. If a 
structural glass is formed via a phase transition it might have 
pure states with similar properties and then the functional 
of Eq. (\ref{BRE}) might describe its transition correctly. 

It is our contention though that there is no phase transition in the 
finite-dimensional p-spin model (in any finite dimension)
and hence discussions of universality
classes are not possible. However, the fact that a large 
correlation length can arise at low temperatures means that 
scaling and universality ideas should  
have a degree of utility and hence that the p-spin model may provide
a qualitative model for structural glasses. According to this possibility 
one could argue
that the more rapid growth than 
Arrhenius for the relaxation time $\tau$ seen in some glasses ( the ``fragile"
glasses) is because the relaxation time grows as
\begin{equation}
\tau=\tau_0 \exp\left[B\xi(T)^{\psi}/kT\right]
\end{equation}
accepting the connection to the droplet model of spin glasses, 
where $\tau_0$ is a microscopic relaxation time,
and the constant $B$ would be expected to be of order $J$. The exponent 
$\psi$ would be that of the zero field Ising spin glass \cite{FH}. The
rapid rise of $\xi(T)$ depicted in Figure \ref{fig1} would then produce a very
non-Arrhenius temperature dependence of the relaxation time $\tau$. On the
other hand some glasses (the ``strong" glasses) do have an 
Arrhenius temperature dependence of their relaxation time. We would
interpret this as arising when $\xi(T)$ at low temperatures 
is not large but instead of order one. It is our contention
that the structural glass problem is similar to the Ising
spin glass in a field, and if that field is large the
correlation length  is small  \cite{BM,Drossel} and the
non-Arrhenius behavior produced by a rapidly increasing 
correlation length will be absent in those circumstances. The validity of this
picture would be greatly strengthened if experimental evidence of 
a growing correlation length at temperatures
around $T_K$ in fragile glasses were to be experimentally observed.

\begin{acknowledgments}
One of us (M.A.M) would like to thank A. J. Bray and A. Cavagna for
numerous discussions.
B.D. was supported by the Deutsche 
Forschungsgemeinschaft (DFG) under Contract
No Dr300-2/1.
\end{acknowledgments}

\end{document}